\documentstyle[seceq,epsf,wrapft,preprint]{ptptex}

\notypesetlogo  
\preprintnumber[3cm]{
UTAP-398/01\\ RESCEU-24/01\\ hep-ph/0110187}

\title{
Determining the Supernova Direction \\by its Neutrinos
}

\author{
Shin'ichiro {\sc Ando}$^{1}$
and Katsuhiko {\sc Sato}$^{1,2}$
}

\inst{
$^1$Department of Physics, School of Science, the University of Tokyo,\\
	Hongo 7-3-1, Bunkyo, Tokyo 113-0033, Japan
\\
$^{2}$Research Center for the Early Universe, \\
	School of Science,		    
	the University of Tokyo,\\
	Hongo 7-3-1, Bunkyo,
     Tokyo 113-0033, Japan}

\recdate{
}

\abst{
Supernova neutrinos, which arrive at Earth earlier than light, allow for the earliest determination of the direction of the supernova.
The topic of this paper is to study how accurately we can determine the supernova direction. 
We simulate supernova neutrino events at the SuperKamiokande detector, using a realistic supernova model and several realistic neutrino oscillation models.
With the results of our simulation, we can restrict the supernova direction to be within a circle of radius $9^\circ$. 
In several neutrino oscillation models, this accuracy is increased to $8^\circ$.
We also discuss the influence of an accident that occurred at the SuperKamiokande detector.
After repair of the detector, using the remaining PMTs, the accuracy becomes about $12^\circ$ for no oscillation. 
}

\begin{document}

\maketitle

\section{Introduction\label{sec:Introduction}}

In order to obtain information concerning core-collapse supernova explosions, which are rare in the galaxy, \cite{rf:Muller} it is very important to observe the light curve of its early phase.
It is expected that a core-collapse supernova explosion in our galaxy will be detected in the future by several neutrino detectors around the world. 
When a supernova collapse occurs, neutrinos produced in the core can escape immediately, because of their very weak interaction with matter.
Contrastingly, photons do not escape until the shock wave travels from the core through the stellar envelope and breaks out of the stellar photosphere. 
For this reason, we can receive the neutrino signal several hours earlier than the light signal. 
(Of course this time delay depends on the size of the envelope. Reference \citen{rf:Shigeyama} contains a simple model of this delay.)
In addition, because electromagnetic signals are obscured by dust in the interstellar space, it is plausible that some supernova explosions cannot be detected without the neutrino signal.
Therefore, if we could determine the direction of a supernova explosion by its neutrinos, many astronomical observations of its early state would be possible.
In fact, a world-wide early supernova alert project is running (SNEWS, or SuperNova Early Warning System). \cite{rf:Scholberg,rf:Habig}

This problem (of determining the direction to a supernova by its neutrinos) has been studied previously in a general manner \cite{rf:Beacom}\tocite{rf:LoSecco}.
There are two methods to deal with this problem.
The first method uses the angular distributions of the neutrino reaction products, which can be correlated with the supernova direction.
Among past works following this approach, Beacom and Vogel in Ref. \citen{rf:Beacom} evaluated the numerical integral to find the centroid of the Gaussian peak of the reaction products' distribution, with a known flat background.
Their result is $\delta \theta \simeq 5^\circ$ for SuperKamiokande, and $\delta \theta \simeq 20^\circ$ for SNO.
The second method is based on triangulation using two or more widely-separated detectors.
However, this technique was shown to be very crude in Ref. \citen{rf:Beacom} ($\delta \cos \theta \simeq 0.5$ for SuperKamiokande and SNO), in contrast to previous optimistic estimates given in, for example, Ref. \citen{rf:Burrows}.  

In this paper, we simulate a supernova explosion in the galactic plane ($D=10$ kpc as used in Ref. \citen{rf:Beacom}, for comparison) and discuss the statistical error in its direction using the first method mentioned above.
This approach, which is based on a more concrete simulation comparing to the past works, is very precise. 
In addition, we also consider the more realistic case, that of ``neutrino oscillation,'' which is supported by solar \cite{rf:Fukuda_1}\tocite{rf:SNO} and atmospheric neutrino \cite{rf:Fukuda_2} data. 
This realistic case was not considered in previous works.
We expect that in the case of neutrino oscillation, the accuracy of the determined supernova direction will be better, since the energy of $\nu_e$'s, which give the largest contribution in determining the direction, is increased due to the conversion $\nu_e\leftrightarrow\nu_{\mu,\tau}$.
(Because $\nu_{\mu,\tau}$'s experience only neutral-current reactions in the supernova and interact with matter more weakly than $\nu_e$'s, $\nu_{\mu,\tau}$'s reach equilibrium deeper in the core than $\nu_e$'s, and their temperatures are higher than those of $\nu_e$'s.)

The construction of this paper is as follows.
In \S \ref{sec:Models}, a realistic supernova model and the neutrino oscillation models used in the simulation are presented. 
Reactions at the SuperKamiokande detector and these cross sections are discussed in \S \ref{sec:SK}.
Based on these models and reactions, we simulated a supernova explosion in our galaxy and events at SuperKamiokande.
The results of these simulations are reported in \S \ref{sec:Results}. 
The details of this simulation and its results concerning the accuracy of the direction are also reported in this section.
Finally we discuss our results in \S \ref{sec:Discussion}.

\section{Supernova model and oscillation parameters \label{sec:Models}}

\begin{figure}[htbp]
\epsfxsize=15.0cm
\centerline{\epsfbox{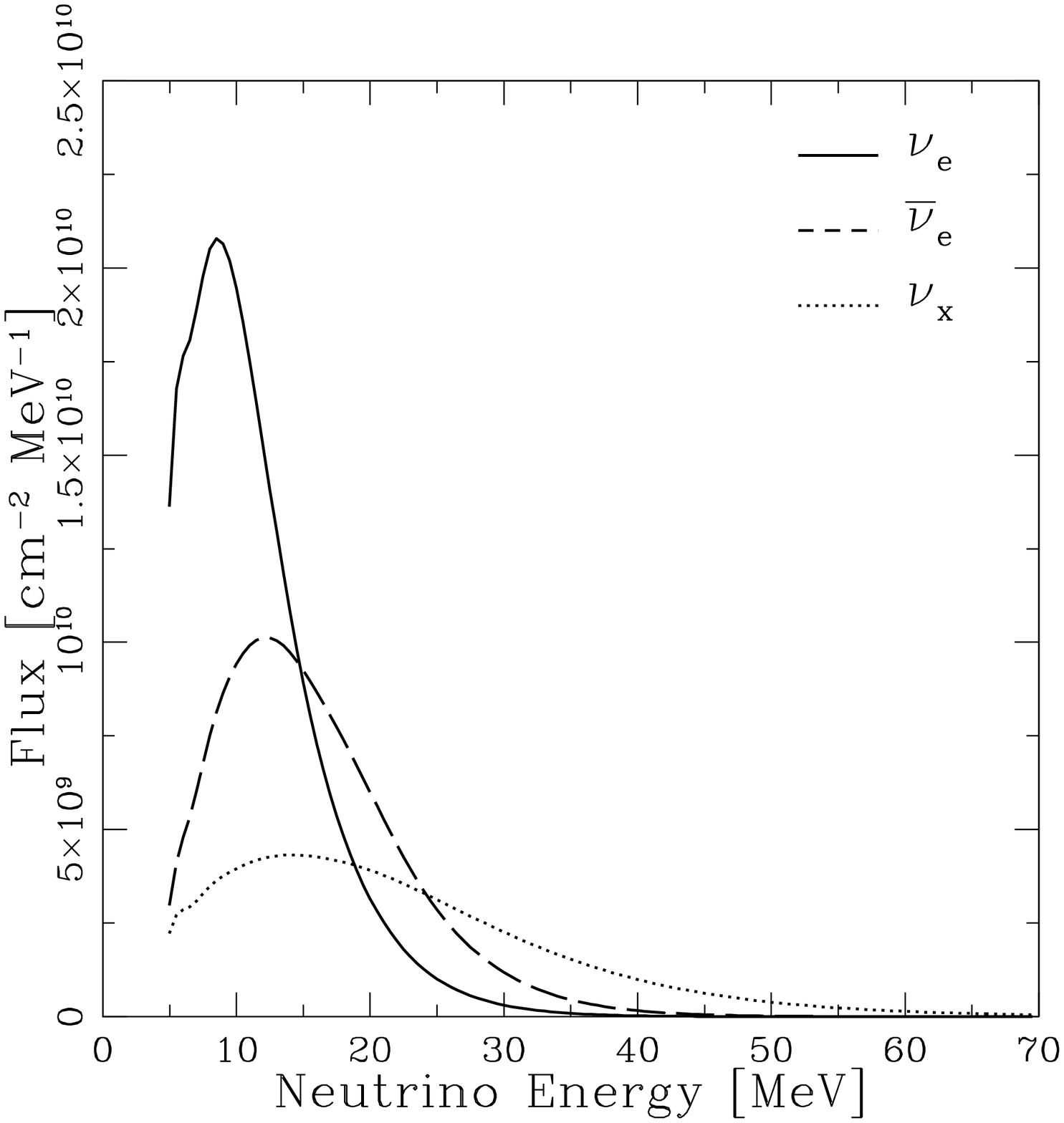}}%
\caption{Energy spectra of $\nu_{e}$, $\bar{\nu}_{e}$ and $\nu_{x}$ of the 
          numerical supernova model used in this paper (, where $\nu_x$ means 
	  $\nu_{\mu,\tau}$ and $\bar{\nu}_{\mu,\tau}$). The solid, dashed and 
          dotted lines are the spectrum of $\nu_{e}$, $\bar{\nu_{e}}$ and 
          $\nu_{x}$, respectively. These spectra are assumed ``no oscillation''\label{fig:flux}}
\end{figure}

We use a realistic model of a collapse-driven supernova proposed by the Lawrence Livermore group. \cite{rf:Wilson} 
Time-integrated energy spectra for the case of no oscillation is shown in Fig. \ref{fig:flux}.
(See Ref. \citen{rf:Totani} for details.)
In this model, however, the radiation of neutrinos is isotropic, despite the fact that supernova progenitors are rotating (which many observations have indicated), and they radiate anisotropic neutrinos. \cite{rf:Shimizu}
For now, we do not consider the effect of rotation for simplicity.


We use four models of neutrino mixing parameters used in Ref. \citen{rf:Takahashi} (see Table \ref{table:parameter}), and the results of the supernova neutrino oscillation given in Ref. \citen{rf:Takahashi} are also used. 
These models are made to agree with the results of the solar and atmospheric neutrinos. \cite{rf:Fukuda_1}\tocite{rf:Fukuda_2}
The expressions ``LMA'' and ``SMA'' indicate the MSW solution of the solar neutrino problem.
A recent SNO observation, \cite{rf:SNO} along with other observations, shows that the LMA solution is better than the SMA solution, and the SMA solution is not allowed below the $3\sigma$ level (see Ref. \citen{rf:Krastev} and references therein).
However, we also consider the SMA solution for comparison. 
The suffices ``-L'' and ``-S'' attached to ``LMA'' and ``SMA'' indicate whether $\theta_{13}$ is large or small, where $\theta_{13}$ is one of three mixing angles of the neutrino mixing matrix.
A large (small) $\theta_{13}$ means that the ``higher resonance'' is adiabatic (nonadiabatic). \cite{rf:Dighe} 
An adiabatic higher resonance enhances the energy of the electron neutrinos, and enhances the event rate of $\nu_e$ scattering, which appears to be strongly correlated with the supernova direction (see below). 
For a review of the MSW effect, see Ref. \citen{rf:Kuo}.


\section{Expected events at SuperKamiokande \label{sec:SK}}

SuperKamiokande (SK) is a water \v{C}herenkov detector with 32,000 tons of pure water located at Kamioka, Japan. 
The relevant interactions of neutrinos with water are as follows:
\begin{eqnarray}
\bar{\nu}_{e} +p & \rightarrow & n + e^+, \quad(\rm{CC}) 
\label{eq:proton}\\
\nu_e + e^- & \rightarrow & \nu_e + e^-, \quad(\rm{CC \quad and \quad NC}) 
\label{eq:electron}\\
\bar{\nu}_{e} + e^- & \rightarrow & \bar{\nu}_{e} + e^- ,
\quad(\rm{CC \quad and \quad NC}) \label{eq:ebar}\\
\nu_{\mu,\tau}(\bar{\nu}_{\mu,\tau}) + e^- & \rightarrow & \nu_{\mu,\tau} (\bar{\nu}_{\mu,\tau})+ e^-, \quad(\rm{NC}) \label{eq:mutau}\\
\nu_e + O & \rightarrow & F + e^-, \quad(\rm{CC}) \label{eq:O1}\\
\bar{\nu}_{e} + O & \rightarrow & N + e^+, \quad(\rm{CC})
\label{eq:O2}
\end{eqnarray}
where CC and NC stand for charged current and neutral current
interactions, respectively. 

\begin{figure}[htbp]
\epsfxsize=15.0cm
\centerline{\epsfbox{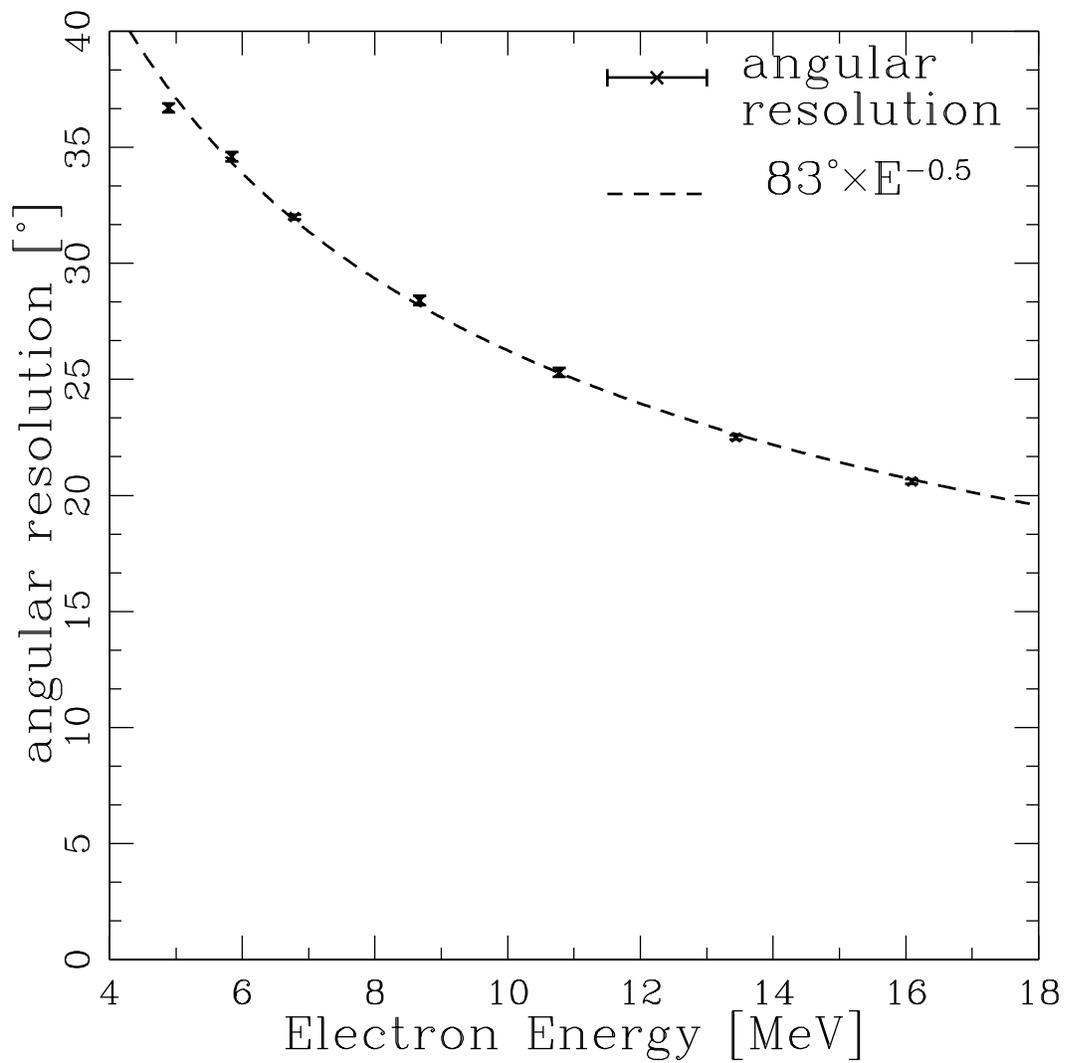}}
\caption{ The angular resolution of the SK detector. 
	The dashed curve represents $83^\circ\times E_e^{-1/2}$,
	where $E_e$ is measured in MeV.
\label{fig:resolution}
}
\end{figure}

The efficiency of the SK detector is 100\% for an electron whose energy is above 5 MeV and 50\% at 4.2 MeV.
(For the energy and angular resolution of SK, we refer to Ref. \citen{rf:Nakahata}.)
The energy resolution is $\sim 15\%$ for an electron with energy 10 MeV.
We display the angular resolution as a function of the recoil electron energy in Fig. \ref{fig:resolution}. 
In this figure, we fitted experimental data to the function $83^\circ E_e^{-0.5}$, where $E_e$ is measured in MeV, which is used in our simulation.

\begin{figure}[htbp]
\epsfxsize=15.0cm
\centerline{\epsfbox{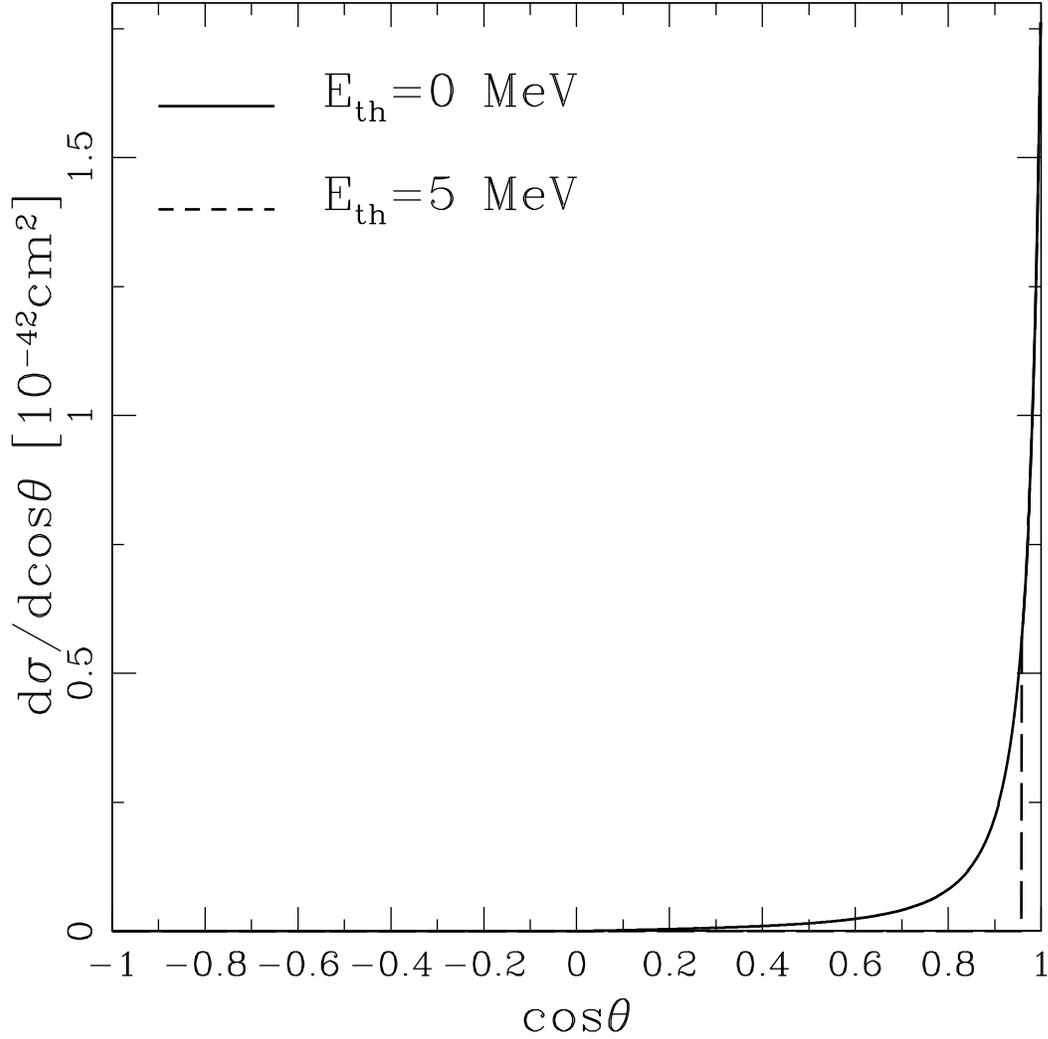}}
\caption{Cross section for $\nu_e+e^-\rightarrow\nu_e+e^-$, with $E_{\nu}=10$ MeV.
	The solid curve represents the cross section without an energy threshold. 
	The dashed curve represents that with an energy threshold of $5$ MeV, 
	and that threshold is of SuperKamiokande detector.
	With the threshold, only the event of $\cos\theta>0.95$ can be seen.
\label{fig:cross_ES}}
\end{figure}

The differential cross section of the electron scattering (\ref{eq:electron})-(\ref{eq:mutau}) (for its derivation, see Ref. \citen{rf:Fukugita}), is given by
\begin{eqnarray}
\frac{d\sigma}{d\cos\theta}&=&\frac{G_F^2T_e^2}{2\pi}\frac{(1+2m_e/T_e)^{3/2}}{1+m_e/E_{\nu}} \nonumber \\
	& & {} \times \left[A+B\left(1-\frac{T_e}{E_{\nu}}\right)^2+C\frac{m_eT_e}{E_{\nu}^2}\right], \label{eq:cross_ES}\\
\cos\theta&=&\frac{E_{\nu}+m_e}{E_{\nu}}\left(\frac{T_e}{T_e+2m_e}\right)^{1/2},
\end{eqnarray}
where $G_F$ is the Fermi constant, $m_e$ is the electron mass, $T_e$ is the electron kinetic energy, $E_{\nu}$ is the neutrino energy, and the coefficients $A, B,$ and $C$ are given in Table \ref{table:coef}.
$\theta$ is the angle between the injected neutrino and the recoil electron (positron).
This differential cross section is highly peaked in the forward direction, as shown in Fig. \ref{fig:cross_ES}. 
With a threshold energy ($5$ MeV), this forward peak is enhanced.

\begin{figure}[htbp]
\epsfxsize=15.0cm
\centerline{\epsfbox{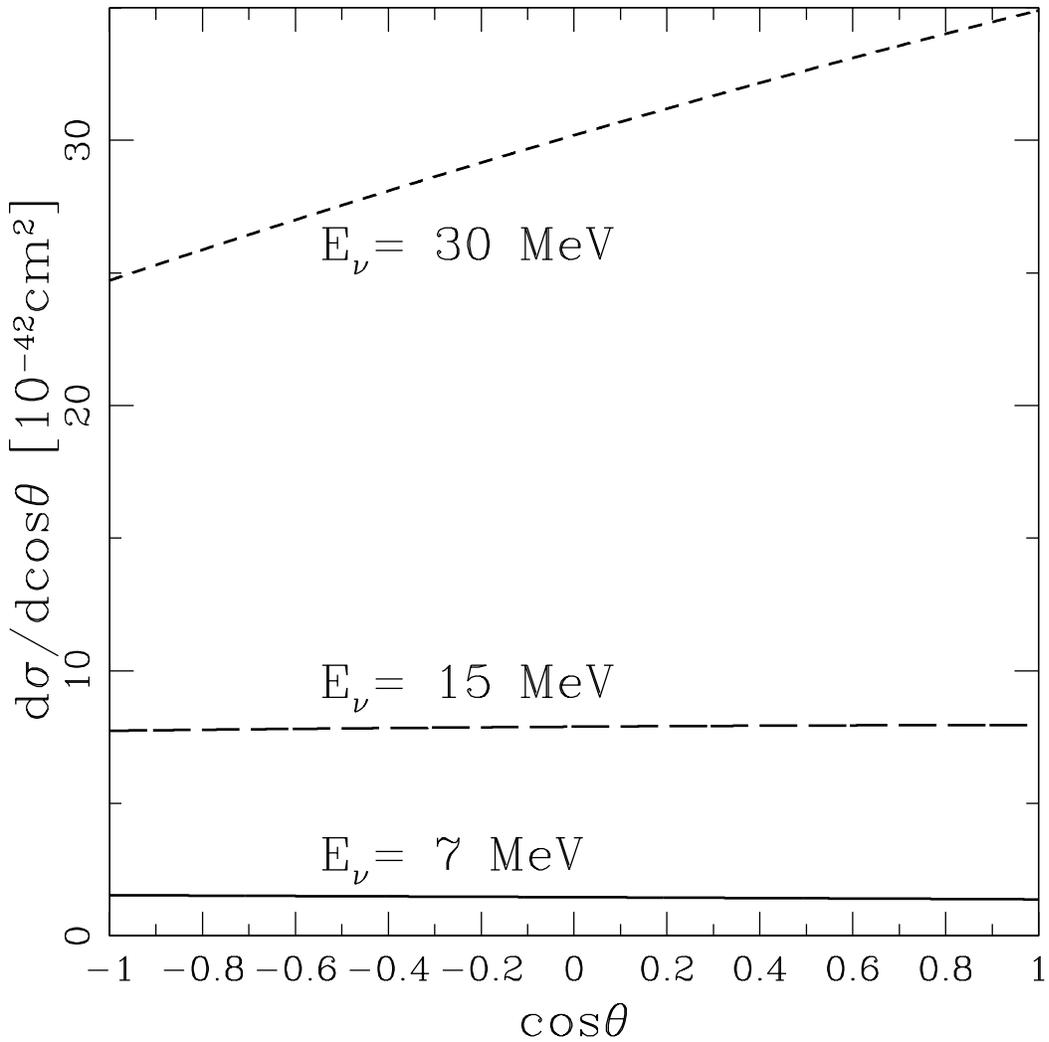}}
\caption{Cross sections for $\bar{\nu}_e+p\rightarrow e^++n$.
	These are almost isotropic. When $E_{\nu}=30$ MeV, a weak forward peak 		can be seen, but this peak is not as sharp as scattering events.
\label{fig:cross_p}}
\end{figure}

The differential cross section of the $\bar{\nu}_{e}p$ CC reaction (\ref{eq:proton}) (see Ref. \citen{rf:Vogel}) is
\begin{eqnarray}
\frac{d\sigma}{d\cos\theta} 
	&=& \frac{\sigma_0}{2}\left[(f^2+3g^2)+(f^2-g^2)\cos\theta -\frac{\Gamma}{M}\right] E_e^{(0)}E_e^{(0)}, \label{eq:cross_p} \\
\Gamma	&=& 2(f+f_2)g[(2E_e^{(0)}+\Delta)(1-\cos\theta)] \nonumber \\
	&+& (f^2+g^2)[\Delta(1+\cos\theta)] \nonumber \\
	&+& (f^2+3g^2)[3(E_e^{(0)}+\Delta)(1-\cos\theta)-\Delta] \nonumber \\
	&+& (f^2-g^2)[3(E_e^{(0)}+\Delta)(1-\cos\theta)-\Delta]\cos\theta,
\end{eqnarray}
where $f=1, g=1.26,$ and $f_2=\mu_p-\mu_n=3.706$ (with $\mu_p, \mu_n$ the magnetic moments of the proton and a neutron, respectively, in units of the nuclear magneton), $M$ is the nucleon mass, and $\Delta$ is the mass difference between a neutron and a proton.
The normalizing constant $\sigma_0$, including the energy-independent inner radiative corrections, is
\begin{eqnarray}
\sigma_0 & = & \frac{G_F^2\cos^2\theta_C}{\pi}(1+\Delta_{inner}^R),
\end{eqnarray}
where $\theta_C$ is the Cabibbo angle ($\cos\theta_C=0.974$) and $\Delta_{inner}^R\simeq 0.024$.
This cross section (\ref{eq:cross_p}) is expressed to first order in $E_{\nu}/M$, and depends on the zeroth-order positron energy $E_e^{(0)}=E_{\nu}-\Delta$.
As shown in Fig. \ref{fig:cross_p}, the differential cross section of the $\bar{\nu}_{e}p$ reaction is almost isotropic, and the $\bar{\nu}_{e}p$ reaction has the largest contribution to the detected events at SK [e.g., at $E_{\nu}=10$ MeV, $\sigma(\bar{\nu}_ep)\simeq 100\sigma(\nu_ee^-)$].
For this reason, it is not easy to see the peak position of events, information about which enables us to determine the supernova direction easily.


The differential cross sections of the reactions with oxygen (\ref{eq:O1}) and (\ref{eq:O2}) are unclear, because of the uncertainty of the the nuclear part. 
These reactions are important if we consider the case with ``neutrino oscillation,'' because the oscillation enhances the energies of $\nu_e$ and $\bar{\nu}_e$, which contribute to the reactions with the oxygen, and therefore enhances the cross sections of these reactions. 
The differential cross sections with oxygen calculated in Ref. \citen{rf:Haxton} are highly peaked in the backward direction when the neutrino energy becomes larger, and therefore the reactions with oxygen seem to be useful for determining the supernova direction.
However, the backward peaks of these reactions are not as sharp as the forward peaks of the electron scattering reactions.
For this reason, we do not consider the reactions with oxygen, because we believe that these reactions influence the result only slightly. 

The number of events at SK is calculated in Ref. \citen{rf:Takahashi} in the case of a supernova explosion at $D = 10$ kpc. 
We give these results in Table \ref{table:event_SK}. (Note that the reactions with oxygen are also shown.)


We cannot expect better accuracy for the direction from SNO observation than from SK observation (see Ref. \citen{rf:Beacom}, for example), and therefore we do not discuss events at SNO.

\section{Simulation and results \label{sec:Results}}

We assume that a supernova explosion occurred at $D=10$ kpc.
To obtain information about the direction of the supernova, we have to set the coordinates at SK.
Here, we set the $z$-axis in the upward direction, and we use spherical coordinates $(\theta, \phi)$ as follows to determine the direction:
\begin{eqnarray}
x = \cos \theta \sin \phi, \\
y = \sin \theta \sin \phi, \\
z = \cos \theta .
\end{eqnarray}
We also assume here that the direction of the supernova is $(\theta, \phi) = (135^\circ, 270^\circ)$. 
In this case, the peak position should be $(\theta, \phi) = (45^\circ, 90^\circ)$. 
(As we see below, this assumption does not influence our final results.)

Recoil electrons and positrons are distributed according to the equations (\ref{eq:cross_ES}) and (\ref{eq:cross_p}).
However, they also experience multiple scattering in water, and we understand this effect as follows.
An electron or a positron of energy $E_e$ that scatters is characterized by a Gaussian probability distribution, whose center value is the original direction of that electron and whose one sigma error is $83^\circ \times (E_e/\mathrm{MeV})^{-1/2}$ (as shown in Fig. \ref{fig:resolution}).
In addition, another factor that effects SK events is the energy resolution of the detector. 
However, we do not have to consider the energy resolution, because it represents the accuracy with which the detector can determine the electron energies, and in our simulation we do not investigate the electron energy distribution.

\begin{figure}[htbp]
\epsfxsize=15.0cm
\centerline{\epsfbox{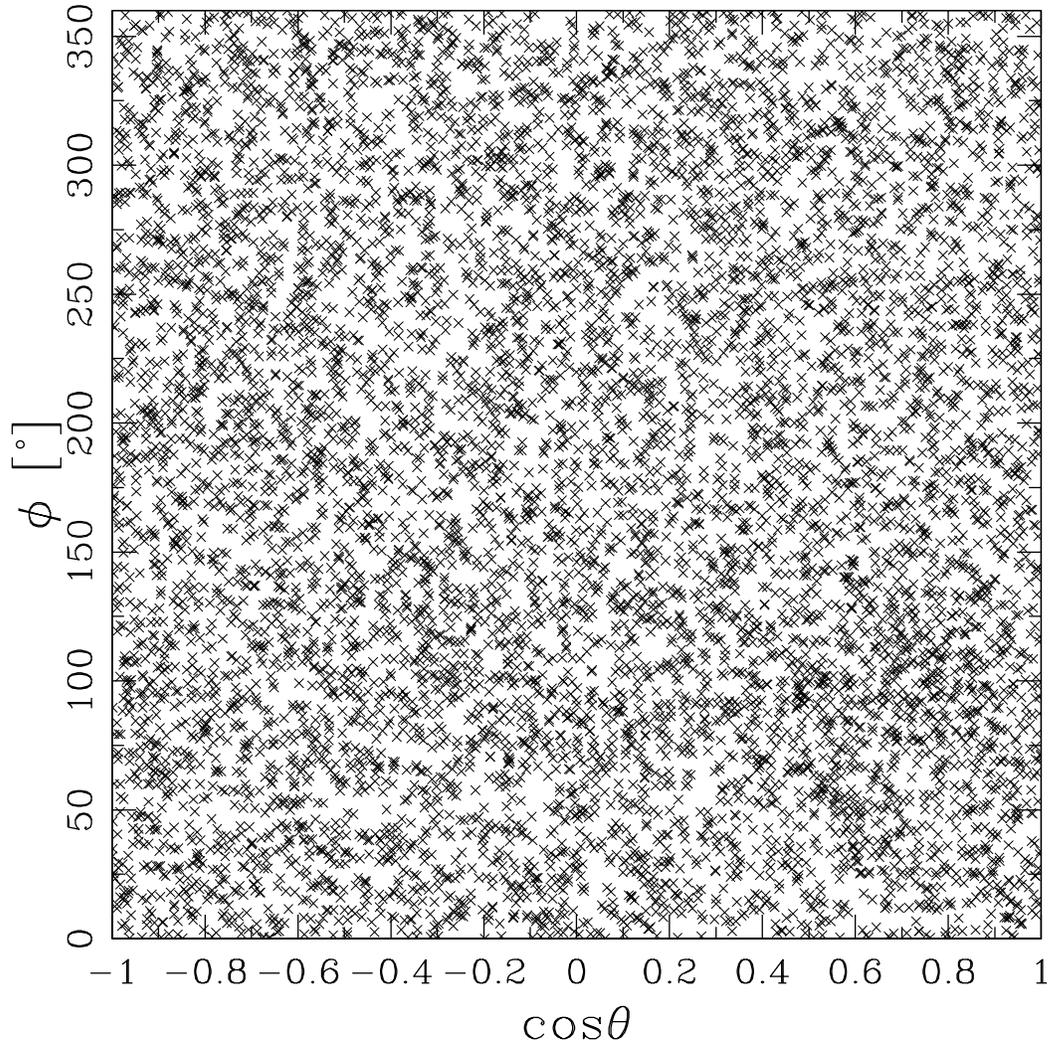}}
\caption{Events at the SuperKamiokande detector. 
	An obscure peak around 
	$(\theta, \phi) \simeq (45^\circ, 90^\circ)$ can be seen.
\label{fig:events}}
\end{figure}

\begin{figure}[htbp]
\epsfxsize=15.0cm
\centerline{\epsfbox{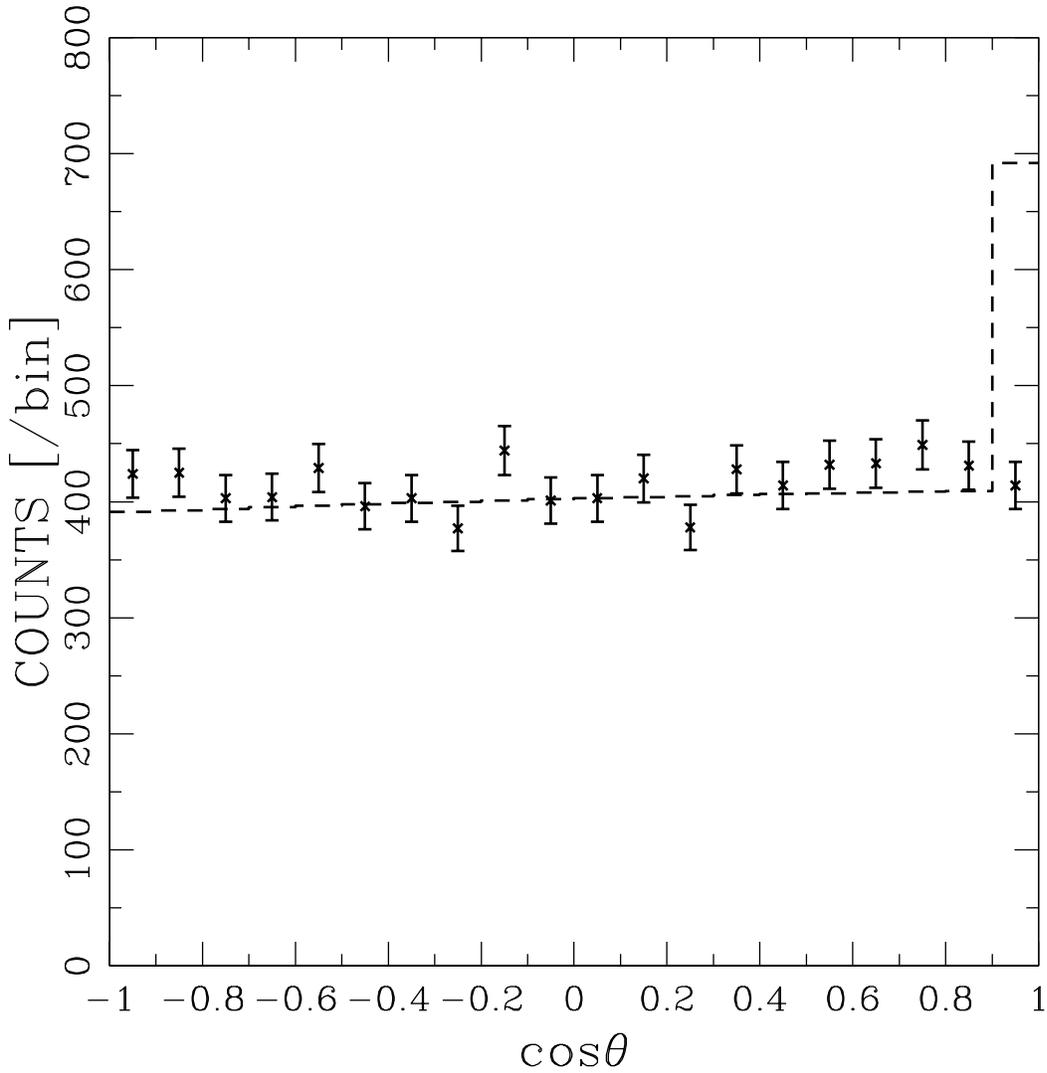}}
\caption{Analyzed figure of the event distribution (Fig. \ref{fig:events}).
	The data points represent the numbers of events in each bin,
	with 20 bins along the $\cos \theta$ direction. 
	The dashed histogram corresponds to the $\theta = 0^\circ$ peaked case, 
	which can easily be calculated theoretically with cross sections.
\label{fig:analyze}}
\end{figure}

\begin{figure}[htbp]
\epsfxsize=15.0cm
\centerline{\epsfbox{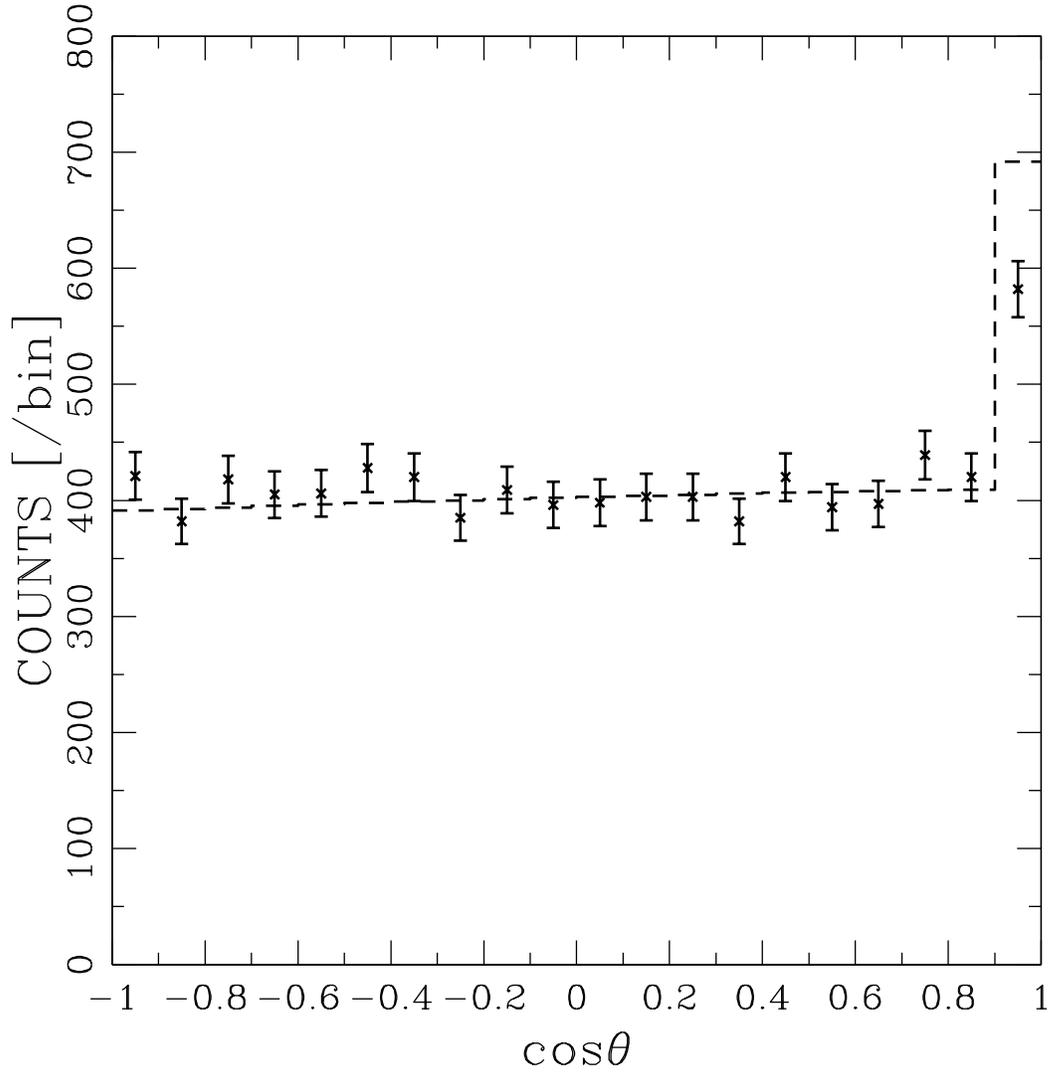}}
\caption{The analyzed figure after the coordinate rotation.
	The rotation angle is determined by a least-square method, so that the
	data points in Fig. \ref{fig:analyze} can be seen as in the case that the peak is located at  
	$\theta = 0^\circ$.
	This best-fitted result is obtained at the angles
	$(\theta_{\rm bestfit}, \phi_{\rm bestfit}) 
	= (47.7^\circ, 80.0^\circ)$.
\label{fig:minanalyze}}
\end{figure}

\begin{figure}[htbp]
\epsfxsize=15.0cm
\centerline{\epsfbox{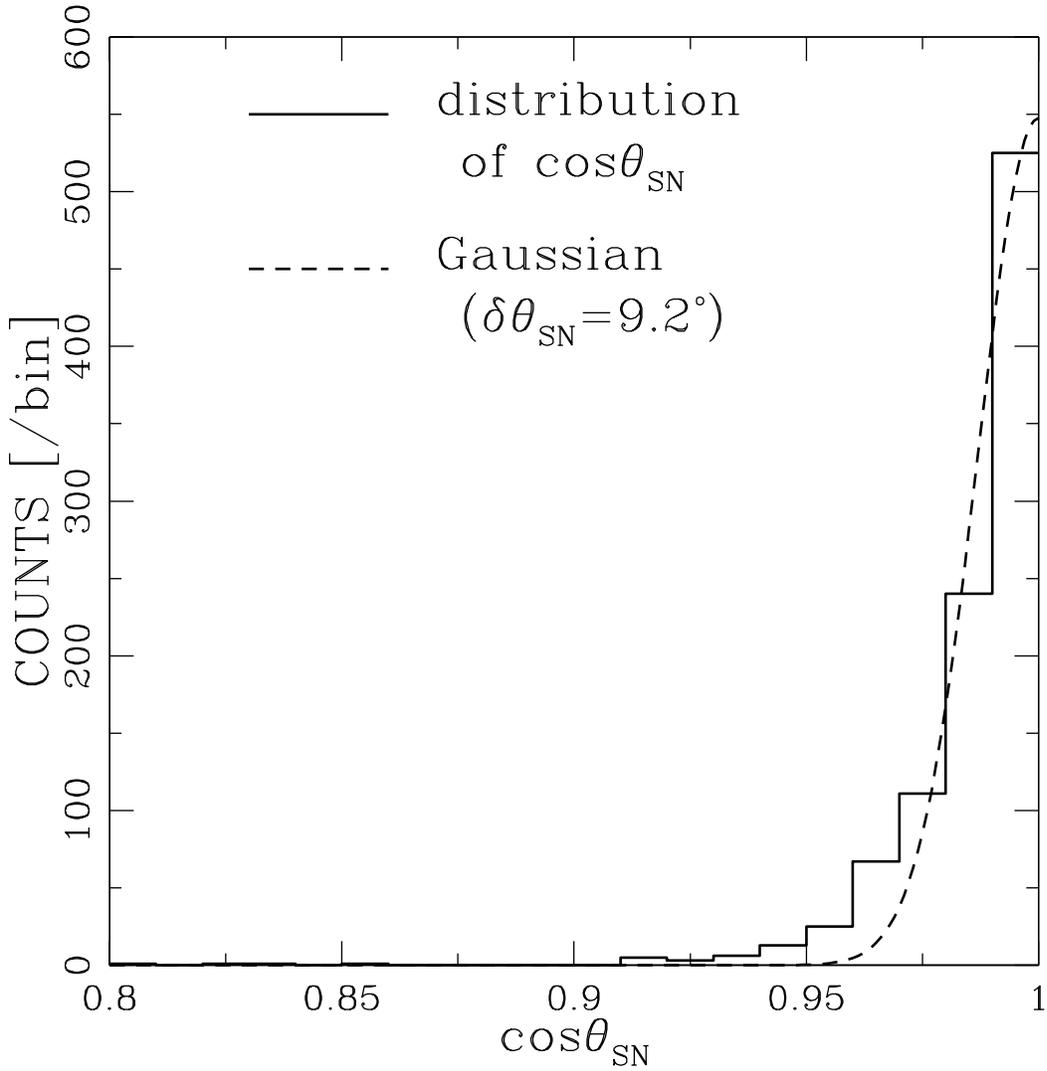}}
\caption{The distribution of $\theta_{\rm SN}$ for 1,000 simulations.
	The values $\theta_{\rm SN}$ are the angles between the 1,000 best-fitted positions 
	derived using a least-square method (see Figs. \ref{fig:analyze},
	\ref{fig:minanalyze}) and the ``true'' peak position we have assumed.
	A Gaussian ($\delta \theta_{\rm SN} = 9.2^\circ$)
	is fitted to the data histogram.
\label{fig:dispersion}}
\end{figure}

First, we consider the case of ``no oscillation.''
We display in Fig. \ref{fig:events} the distribution of the events for this case.
There seems to be an obscure peak around $(\theta,\phi) = (45^\circ,90^\circ)$.
To analyze this result systematically, we divide the $\cos \theta$ direction into 20 bins and sum up in each bin.
Figure \ref{fig:analyze} is the result of this operation.
The dashed curve in Fig. \ref{fig:analyze} corresponds to the $\theta = 0^\circ$ peaked case, which can easily be calculated theoretically with cross sections.
Then, we rotate the coordinates so that the events are seen as in the case that the peak is located at $\theta = 0^\circ$ using a least-square method.
The best-fitted result is shown in Fig. \ref{fig:minanalyze}.
This result is obtained at angles  $(\theta_{\rm bestfit}, \phi_{\rm bestfit}) = (47.7^\circ, 80.0^\circ)$. 
This result is near the ``true'' values, $(45^\circ,90^\circ)$, but this simulation alone is not sufficient to estimate the errors.
We, carried out the simulation 1,000 times under the same conditions and obtained a distribution of the best-fitted coordinate rotation angles.
We denote by $\theta_{\rm SN}$ the angle between these best-fitted points $(\theta_{\rm bestfit},\phi_{\rm bestfit})$ and the ``true'' point $(45^\circ,90^\circ)$, and we measure these discrepancies $\theta_{\rm SN}$.
[We should note that when we analyze using $\theta_{\rm SN}$, the information regarding the original peak position, or $(45^\circ,90^\circ)$, is lost, and our analysis from this point does not depend on where the peak position is.]
The solid histogram in Fig. \ref{fig:dispersion} is the distribution of the values $\cos\theta_{\rm SN}$, and we fitted this result to a Gaussian (dashed curve).
The one sigma error of this Gaussian is $\delta \cos\theta_{\rm SN}=0.0130$.
By using the simple formula $\delta\cos\theta_{\rm SN}\simeq \delta\theta_{\rm SN}^2/2$, we obtain $\delta \theta_{\rm SN}=9.2^\circ$.
Then in this method we can determine the supernova direction with an accuracy of $9.2^\circ$.


We adapted the same simulations to the other four ``neutrino oscillation'' cases.
We summarize these results in Table \ref{table:sigma}.  


\section{Discussion \label{sec:Discussion}}

As shown in Table \ref{table:sigma}, we can determine the supernova direction within $\sim10^\circ$.
In the five models with which we have dealt in this paper, ``LMA-L'' and ``SMA-L'' are the best, because the electron scattering events are more prominent for them, as shown in Table \ref{table:event_SK}.

Now we give further discussion of three points below, where we assume the case of ``no oscillation.''

\subsection{Oxygen events}
As discussed above and shown in Table \ref{table:event_SK}, we must consider the oxygen events, especially in the case of ``neutrino oscillation.''
We do not expect that these effects will enable us to determine the direction more accurately than in the no oscillation case.
Rather, because their reaction cross sections are unclear, they may complicate our analyses by acting as noise. 

If we would like to determine the supernova direction without the oxygen, whose effect is very complicated, one method is to include an ``energy cut-off.''
For example, we only consider electrons whose energies are less than 15 MeV, so that oxygen events should have little contribution. (However, we cannot be sure that an energy cut-off of 15 MeV is sufficient to accomplish this.)  
With this energy cut-off, we carried out simulations in the same manner.
For the result obtained in this case, the accuracy is rather worse, $\delta\theta_{\rm SN} \sim 13^\circ$.
This is because lower energy electrons (positrons) are scattered in the detector, and this makes the angular resolution of these electrons (positrons) worse, as shown in Fig. \ref{fig:resolution}.

\subsection{Dependence on distance}

The simulation and its results given above are based on the assumption that the supernova exploded at a distance $D=10$ kpc.
Since the event number at the earth falls off as the distance $D$ squared, $N\propto D^{-2}$, and the accuracy $\delta\theta_{\rm SN}$ is proportional to $1/\sqrt{N}$, we expect that the accuracy is proportional to the distance, or, 
\begin{equation}
\delta\theta_{\rm SN} = 9.2^\circ \left(\frac{D}{10 ~\mathrm{kpc}}\right).
\label{eq:distance}
\end{equation}
Actually, we carried out simulations assuming other distance ($D=5,7.5,12.5,15,17.5,20$ kpc) and obtained approximately the same results as above equation.

\subsection{After the accident of SuperKamiokande}
On November 12th, 2001, an accident occurred at the SuperKamiokande detector, and a significant part (about $60\%$ of the PMTs) of the detector was damaged.
The Superkamiokande detector is expected to restart within a year or so, by rearranging the PMTs, whose number density is reduced by a half, using the existing resources, and using the same volume of water as before.
After this repair, the effect of the accident on its performance is expected not to be serious for supernova neutrinos, because the fiducial volume will not change, and the threshold energy change (from 5 MeV to about 7-8 MeV) \cite{rf:Totsuka} influences the event number very little, since the low energy event number is small. 
(For example, the total event number of electron scattering and the inverse $\beta$ decay reaction is 8,318 for a threshold of 5 MeV and is 8,165 for a threshold of 8 MeV, based on the calculation of Ref. \citen{rf:Takahashi}.)
The energy resolution will be rather large, but this is not a serious problem, since it is proportional to square root of the number of PMTs. \cite{rf:Totsuka}

For analysis of the direction, the detector performance is not changed significantly by the change in the detector.
The reason for this is as follows.
First, since the fiducial volume does not change, the total event number decreases, only because of the higher energy threshold.
This effect is very small, as shown above.
Second, the angular resolution of the detector, which is most important in determining the accuracy, is affected little because the higher energy threshold cuts low energy electrons (positrons), which mainly obscure the peak.
For these reasons, the angular resolution is not strongly affected, and therefore neither is the accuracy in determining the supernova direction. 

Actually, our simulation with threshold 8 MeV gives $\delta\theta_{\rm SN}=11.9^\circ$.
This difference comes mainly from the relative smallness of the electron scattering events.
But this effect cancels with the higher angular resolution, discussed as the second reason above, and results in the small difference $2.7^\circ$.

We expect that the SuperKamiokande detector will be repaired soon.

\section*{Acknowledgements}
We would like to thank the staff at SuperKamiokande, including Y. Totsuka, Y. Suzuki, and Y. Fukuda, for useful discussions, and also we would like to thank K. Takahashi for preparing the numerical data for the neutrino oscillation and for useful discussions. 
This work was supported in part by
Grants-in-Aid for Scientific Research provided by the Ministry of Education,
Science and Culture of Japan through Research Grant No. 07CE2002.

\clearpage

\clearpage

\begin{table}
\caption{Sets of mixing parameters used in the calculation. Here, $\theta_{ij}$ represents
	the mixing angle of the neutrino mixing matrix and $\Delta m_{ij}^2$
	represents the squared mass difference between the $i$-th and $j$-th 
	mass eigenstate of the neutrinos.
	\label{table:parameter}}
\begin{center}
\begin{tabular}{ccccccccc}\hline \hline
model  & $\sin^{2} 2 \theta_{12}$ & $\sin^{2} 2 \theta_{23}$ & $\sin^{2} 2 \theta_{13}$ 
& $\Delta m_{12}^{2}({\rm eV}^{2})$  & $\Delta m_{13}^{2}({\rm eV}^{2})$ 
& $\nu_{\odot}$ problem \\ \hline 
LMA-L &  0.87  & 1.0 & 0.043 & $7.0 \times 10^{-5}$ & $3.2 \times 10^{-3}$ & LMA \\ 
LMA-S &  0.87  & 1.0 & $1.0 \times 10^{-6}$ & $7.0 \times 10^{-5}$ & $3.2 \times 10^{-3}$ 
& LMA \\   
SMA-L &  $5.0 \times 10^{-3}$  & 1.0 & 0.043  & $6.0 \times 10^{-6}$ & $3.2 \times 10^{-3}$ 
& SMA \\ 
SMA-S &  $5.0 \times 10^{-3}$  & 1.0 & $1.0 \times 10^{-6}$ & $6.0 \times 10^{-6}$ & $3.2 \times 10^{-3}$
& SMA \\ \hline
\end{tabular}
\end{center}
\end{table}

\begin{table}[htbp]
\caption{Coefficients for the cross section of $\nu e^- \rightarrow \nu e^-$.
	Here, $g_V=2\sin^2\theta_W-\frac{1}{2}$, and $g_A=-\frac{1}{2},$ where $\theta_W$ is the Weinberg angle ($\sin^2 \theta_W = 0.23$).\label{table:coef}}
\begin{center}
\begin{tabular}{c|ccc} \hline \hline
	coefficient & $A$ & $B$ & $C$ \\
	\hline 
	$\nu_e+e^-\rightarrow\nu_e+e^-$ & $(g_V+g_A+2)^2$ & $(g_V-g_A)^2$ & $(g_A+1)^2-(g_V+1)^2$ \\
	$\bar{\nu}_e+e^-\rightarrow\bar{\nu}_e+e^-$ & $(g_V-g_A)^2$ & $(g_V+g_A+2)^2$ & $(g_A+1)^2-(g_V+1)^2$\\
	$\nu_{\mu,\tau}+e^-\rightarrow\nu_{\mu,\tau}+e^-$ & $(g_V+g_A)^2$ & $(g_V-g_A)^2$ & $g_A^2-g_V^2$\\
	$\bar{\nu}_{\mu,\tau}+e^-\rightarrow\bar{\nu}_{\mu,\tau}+e^-$ & $(g_V-g_A)^2$ & $(g_V+g_A)^2$ & $g_A^2-g_V^2$\\  \hline
\end{tabular}
\end{center}
\end{table}

\begin{table}[htbp]
\caption{Number of events at SuperKamiokande.
	\label{table:event_SK}}
\begin{center}
\begin{tabular}{c|ccccc} \hline \hline
	model & LMA-L & LMA-S & SMA-L & SMA-S 
              & no osc\\
	\hline 
	$\bar{\nu}_{e}p$ & 9459 & 9427 & 8101 & 7967 & 8036 \\ 
	$\nu_ee^{-}$ & 186 & 115 & 189 & 131 & 132 \\ 
	$\bar{\nu}_{e}e^{-}$& 46 & 46 & 41 & 42 & 42 \\ 
	$\nu_{\mu}e^{-}$ & 25 & 26 & 25 & 30 & 30 \\ 
	$\bar{\nu}_{\mu}e^{-}$ & 24 & 23 & 24 & 24 & 24 \\
	$\nu_{\tau}e^{-}$ & 25 & 26 & 25 & 30 & 30 \\ 
	$\bar{\nu}_{\tau}e^{-}$& 24 & 23 & 24 & 24 & 24 \\ 
	$O\nu_e$ & 297 & 214 & 297 & 108 & 31 \\
	$O\bar{\nu}_e$ & 160 & 158 & 95 & 92 & 92 \\ \hline
	total & 10245 & 10114 & 8822 & 8447 & 8441 \\  \hline
\end{tabular} 
\end{center}
\end{table}

\begin{table}[htbp]
\caption{Errors of best-fitted angles of the coordinate rotation in degrees.
	\label{table:sigma}}
\begin{center}
\begin{tabular}{cc} \hline \hline
	model & $\delta \theta_{\mathrm{SN}}$ \\ \hline 
	LMA-L & 8.1 \\
	LMA-S & 10.7 \\
	SMA-L & 8.1 \\
	SMA-S & 9.3 \\
	no osci & 9.2 \\ \hline
\end{tabular}
\end{center}
\end{table}

\end{document}